\documentclass[aps,prb,twocolumn,showpacs,floatfix]{revtex4}

\usepackage{amsmath,amssymb,dsfont,graphicx,psfrag}
\usepackage{color,epsfig,rotate}

\newcommand{\DTN}{NiCl$_{2}$-4SC(NH$_{2}$)$_{2}$}
\newcommand{\CuBr}{(C$_{5}$H$_{12}$N)$_{2}$CuBr$_{4}$}
\newcommand{\TlCu}{TlCuCl$_{3}$}
\begin{document}

\title{Low temperature thermodynamic properties near the field-induced quantum critical point in \DTN}

\author{Franziska Weickert$^{1,2}$, Robert K\"uchler$^{1}$, Alexander Steppke$^{1}$, Luis Pedrero$^{1}$, Michael Nicklas$^{1}$, Manuel Brando$^{1}$, Frank Steglich$^{1}$, Marcelo Jaime$^{2}$, Vivien S. Zapf$^{2}$, Armando Paduan-Filho$^{3}$, Khaled A. Al-Hassanieh$^{4}$, Cristian D. Batista$^{4}$, Pinaki Sengupta$^{5}$}

\affiliation{$^{1}$Max-Planck-Institut f\"ur Chemische Physik fester
Stoffe, 01187 Dresden, Germany\\
$^{2}$Los Alamos National Laboratory, MPA-CMMS, Los Alamos, NM 87545, USA\\
$^{3}$Instituto de Fisica, Universidade de S\~{a}o Paulo, S\~{a}o Paulo, Brazil\\
$^{4}$Los Alamos National Laboratory, Theory Devision T4, Los Alamos, NM 87545, USA\\
$^{5}$School of Physical and Mathematical Sciences, Nanyang Technological University, Singapore 637371}

\date{\today}
\begin{abstract}
We present a comprehensive experimental and theoretical investigation of the thermodynamic properties: specific heat, magnetization and thermal expansion in the vicinity of the field-induced quantum critical point (QCP)
around the lower critical field $H_{c1} \approx 2$\,T in \DTN . A $T^{3/2}$ behavior in the specific heat and magnetization is observed at very low temperatures at $H=H_{c1}$ that is consistent with the universality class of Bose-Einstein condensation of magnons. The temperature dependence of the thermal expansion coefficient at $H_{c1}$ shows  minor deviations from the expected $T^{1/2}$ behavior. Our experimental study is complemented by analytical calculations and Quantum Monte Carlo simulations, which reproduce nicely the measured quantities. We analyze the thermal and the magnetic Gr\"{u}neisen parameters that are ideal quantities to identify QCPs. Both parameters diverge at $H_{c1}$ with the expected $T^{-1}$ power law. By using the Ehrenfest relations at the second order phase transition, we are able to estimate the pressure dependencies of the characteristic temperature and field scales.
\end{abstract}
\maketitle
\section{Introduction}
\label{sec:intro}

Bose-Einstein condensation (BEC) has triggered great interest in the last years and was found in a variety of complex many body systems, such as cold atoms, superfluid Helium or superconductors.
By exploiting the Matusbara-Matsuda mapping of $S=1/2$ spins into hard core bosons, \cite{Matsubara56} Batyev showed  that the field induced phase transition between canted XY antiferromagnetic (AFM) ordering and the fully polarized state can also be described as a BEC. \cite{Batyev84}
This useful mapping between magnetic systems and dilute gases of bosons can be extended to higher spin values \cite{Batista01,Batista04}
and it has been successfully exploited first on \TlCu \cite{Nikuni00} and other quantum magnets.\cite{Giamarchi08,Zapf12}

One material investigated recently, is \DTN ,\cite{PaduanFilho81} also known as dichlorotetrakisthiourea-nickel (DTN). It has a body-centered tetragonal crystal structure with chains of Ni-Cl-Cl-Ni atoms arranged along the crystallographic $c$ direction. DTN enters the XY-AFM ordered state between moderate fields of 2\,T and 12.5\,T, if the magnetic field $H$ is applied along $c$. The magnetic atom in DTN is Ni$^{2+}$ carrying a spin $S=1$ due to an almost completely quenched orbital momentum. The Hamiltonian for DTN can be written as

\begin{equation}
\mathcal{H} =  \sum_{{\bf r} \nu } J_{\nu}  {\bf S}_{\bf r} \cdot {\bf S}_{\bf r +{\bf e}_{\nu}} + \sum_{\bf r} [D (S^{z}_{\bf r})^{2} - g\mu_{B}H S^{z}_{\bf r}],
\label{hamilton}
\end{equation}
where $\nu=\{a,b,c\}$ and ${\bf e}_{\nu}$ is the relative vector between nearest-neighbors along the $\nu$-direction. The magnitude of the dominant single-ion anisotropy $D$ is 8.9\,K.\cite{Zvyagin07} The AFM exchange interactions between neighboring spins are $J_{c}$= 2.2\,K along the chains, and about 10 times smaller, $J_{ab}=$\,0.18\,K, in the $ab$-plane.  The last  Zeeman term in equation\,(\ref{hamilton}) is originated from the applied magnetic field $H$ and the quantization $z$-axis is chosen along the field direction. The gyromagnetic factor $g$ parallel to the $c$-axis was estimated to be 2.26 by ESR experiments.\cite{Zvyagin07} Equation\,(\ref{hamilton}) is only a minimal Hamiltonian
for describing the magnetic properties of DTN. Further contributions, such as dipolar interactions, which break the U(1) symmetry of global rotations along the spin $z$-axis are small, but they become relevant at  very low temperatures. Therefore, the critical exponents characteristic of a BEC quantum critical point (QCP) can only be observed, if the U(1) symmetry breaking terms are at least one order of magnitude smaller than $J_{ab}$. At low enough temperatures one should observe a crossover from the behavior characteristic for  BEC-QCPs to the one expected for an Ising-like QCP.   Investigations of the exact shape of the phase boundary close to $H_{c1}$ and $H_{c2}$ down to 1\,mK by detailed AC susceptibility measurements evidenced the universality class of a BEC in DTN.\cite{Yin08} Up to date, this is the solely experimental observation consistent with a field induced BEC-QCP  in this material.


The universality class of the QCP can also be determined by measuring the exponents  for the power-law dependencies of different thermodynamic quantities as a function of temperature. Table\,\ref{exponents} shows the expected exponents for BEC and Ising-like QCPs in 2 and 3 dimensions $d$.\cite{Zapf12} It is important to note, that $d=3$ is the upper critical dimension for the Ising-like QCP ($D=d+z=4$), where $z=1$ is the dynamical exponent and $D$ the effective dimensionality. Therefore, one should expect further logarithmic corrections to the power law behaviors listed in Table\,\ref{exponents}.
\begin{table}
\begin{center}
\begin{tabular}{|c|c|c|} \hline
                             & XY-AFM order  & Ising    \\
  \hline \hline
  $M(H_c,T)$               & $T^{d/2}$       &     $T^{2}$     \\
  $\frac{\Delta L}{L} (H_c,T)$   & $T^{d/2}$         & $T^{2}$  \\
  $\alpha (H_c,T)$               & $T^{d/2-1}$       & $T$      \\
  $C(H_c,T)$                 & $T^{d/2}$         & $T^{3}$  \\
  \hline
\end{tabular}
\end{center}
\caption{Temperature dependencies of the thermodynamic quantities: magnetization $M(T)$, thermal expansion $\frac{\Delta L}{L},\,\alpha(T)$ and specific heat $C(T)$
at the field induced QCP. The variable $d$ denotes the spatial dimensionality  of the system. The exponents of the Ising-like QCP are given for $d=3$. \label{exponents}}
\end{table}
In this work, an extensive study of the magnetization $M(H,T)$, specific heat $C(H,T)$, and thermal expansion $\alpha_{V}(H,T)$ close to the critical field $H_{c1}$ gives further strong evidence that DTN belongs to the universality class of BEC.

It has been shown recently,\cite{Zhu03} that  QCPs can be detected by measuring the  divergence of the thermal Gr\"{u}neisen parameter,
\begin{gather} \label{gamma_th}
\Gamma_{th}= \frac{\alpha_{V}}{C},
\end{gather}
for pressure tuning, and the magnetic Gr\"{u}neisen parameter,
\begin{gather}
\Gamma_{mag}= -\frac{\partial M/\partial T}{C},
\end{gather}
for a magnetic field tuned QCP. Both Gr\"{u}neisen parameters diverge at the QCP like $\Gamma \propto T^{-1/\nu z}$, where $\nu$ is the
critical exponent that relates the correlation length with the driving parameter of the quantum phase transition (magnetic field in the case of DTN).
The effective dimension $D=d+z$ is higher or equal to 4 with $z=1$ for the Ising-type and $z=2$ for the BEC-type QCP, and we get $\nu=1/2$ for both cases.
Therefore, $\Gamma \propto T^{-1}$ is expected for a BEC-QCP and
$\Gamma \propto T^{-2}$ for an Ising-like QCP in agreement with the power laws listed in Table\,\ref{exponents}.

Typical for QCPs is the occurrence of local maxima in the entropy due to enhanced quantum critical fluctuations. This implies a sign change of the thermal expansion coefficient,\cite{Garst05} which is linked to the entropy via the Maxwell relation $\alpha_{V}= -\partial S/\partial p$. In the past, the concept of the diverging Gr\"{u}neisen parameter was used successfully to identify and characterize not only well understood magnetic QCPs, but also other more puzzling  QCPs in intermetallic compounds.\cite{Kuechler03}


So far, dilatometric properties were used to  investigate the quantum critical behavior of only a few quantum magnets. The $d=3$ coupled spin-dimer system \TlCu \cite{Takatsu97} and the
quasi-one-dimensional spin-ladder compound \CuBr \cite{Patyal90} are two rare examples. These systems show field induced phase transitions at low temperatures,\cite{Oosawa99,Nikuni00,Watson01}
however, investigations of the thermal expansion coefficient $\alpha$ and the Gr\"{u}neisen ratio $\Gamma_{th}$ show significant deviations from the expected behavior in both cases. Dilatometric experiments on \TlCu\ reveal that while the thermal Gr\"{u}neisen parameter diverges with the expected power law $1/T$, the individual quantities specific heat $C$ and thermal expansion $\alpha$ fail to follow the predictions.\cite{Lorenz07} In \CuBr , the thermal expansion coefficient $\alpha_{c}$ along the crystallographic $c$ direction shows a weak indication of $1/\sqrt{T}$ divergency and a clear sign change at the lower and upper critical field $H_{c1/c2}$, but a detailed discussion of $\Gamma_{th}$ is missing.\cite{Lorenz08} The present study of the thermal and magnetic Gr\"{u}neisen ratios shows that DTN is an excellent candidate to close this gap of knowledge about dilatometric properties of insulating quantum critical materials.

The paper is organized as follows: in Section \ref{sec:meth}, we describe the experimental techniques that we used in static magnetic fields to measure the specific heat, thermal expansion and magnetization up to 15\,T, the approximated model used for the analytical calculations, and the quantum Monte Carlo (QMC) simulations of the thermodynamic quantities. Section \ref{sec:exp} contains a detailed description of the experimental and theoretical results. We continue in Section \ref{sec:dis} with the comparison between experiment and theory and the analysis of the anomalies at the phase boundary via the Ehrenfest relations. Section \ref{sec:sum} summarizes the most important results of our study.

\section{Methods}
\label{sec:meth}

The preparation of high-quality single-crystals is explained elsewhere.\cite{PaduanFilho04a} All experiments were conducted, partly down to 30\,mK, inside commercial available dilution refrigerators,
furbished with superconducting (SC) magnets with maximum fields of up to 15\,T. We measured the magnetization with a high resolution Faraday magnetometer.\cite{Sakakibara94} The thermal expansion and magnetostriction experiments were carried out with a high precision capacitive dilatometer\cite{Pott83} made of CuBe. The dilatometer can be rotated by 90$^{\circ}$ in order to measure the length change not only parallel but also perpendicular to the applied magnetic field. The specific heat was measured with the compensated heat pulse technique\cite{Wilhelm04} and the data were confirmed by experiments using the dual-slope method\cite{Riegel86} on the same sample platform. The precise match between both sets of experimental data is remarkable. Additionally, we used the specific-heat setup to perform magnetocaloric effect (MCE) measurements for a precise estimate of the critical field $H_{c1}$.

The analytical calculations of the various thermodynamic properties were based on the usual expansion in the gas parameter or ratio between the scattering amplitude and the average inter-particle distance, $\rho^{-1/3}$. \cite{Abrikosov}  For this purpose we mapped the $S^z=1$ magnetic excitations of the low field paramagnetic (PM) state into hard core bosons, where the $z$-component
of the magnetization density in the original model, $\langle S^z_{\bf r} \rangle$, is mapped into the particle density $\rho$. Here we neglect the contribution to the magnetization of the $S^z=-1$ modes, because we are assuming that $H$ is close to $H_{c1}$ and $T \ll \Delta$, with $\Delta \simeq 3K$ being the $H=0$ spin gap of DTN. We use the expression derived in
Ref.\,[\onlinecite{Kohama11}] for the single-particle dispersion , $\omega_{\bf k}=\omega^0_{\bf k} - g \mu_B H$ with
\begin{equation}
\omega^0_{\bf k}=\sqrt{\mu^2+2 \mu s^2 \epsilon_{\bf k}} .
\end{equation}
The parameters $s^2$ and $\mu$ are given by the following expressions:
\begin{eqnarray}
s^2 = 2 - \frac{1}{N} \sum_{{\bf k}} \frac{\mu + s^2 \epsilon_{\bf k}}{\omega^0_{\bf k} },
\;\;\;
D = \mu + \frac{\mu}{N} \sum_{{\bf k}} \frac{\epsilon_{\bf k}}{\omega^0_{\bf k} }\, .
\label{smu}
\end{eqnarray}
By using the Hamiltonian parameters for DTN estimated in Ref.\,[\onlinecite{Zvyagin07}] the resulting values are $s^2=0.92$ and $\mu=10.3$\,K.
The effective repulsion between bosons in the long wavelength limit, $v_0= \Gamma_{\bf 0} ({\bf Q}, {\bf Q})$ with ${\bf Q}$ being the ordering wave-vector, results from summing the ladder diagrams for the bare interaction vertex $V_{\bf q}$ \cite{Abrikosov}:
\begin{equation}
\Gamma_{\bf q} ({\bf k}, {\bf k}') = V_{\bf q} - \int_{-\pi}^{\pi} \frac{dp^3}{8 \pi^3}  \frac{\Gamma_{\bf p} ({\bf k}, {\bf k}')}{\omega_{\bf k+p}+\omega_{\bf k'-p}},
\end{equation}
where $V_{\bf q} = U + 2 J_c \cos{q_z} + 2 J_{ab}  (\cos{q_x} + \cos{q_y} ) $ for DTN and $U \to \infty $ is included to enforce the hard core constraint. The effective Hamiltonian in the long
wavelength limit $|{\bf k}- {\bf Q}| \ll 1 $ is given by
\begin{equation}
{\cal H}_{\rm eff} = \sum_{\bf k} (\epsilon_{\bf k} - \mu)  a^{\dagger}_{\bf k}  a^{\;}_{\bf k} + \frac{v_0}{2N}  \sum_{{\bf k}, {\bf k'}, {\bf q}}
a^{\dagger}_{\bf k+q}  a^{\dagger}_{\bf k'-q} a^{\;}_{\bf k}  a^{\;}_{\bf k'},
\end{equation}
where $N$ is the total number of lattice sites and  the operator $a^{\dagger}_{\bf k}$ ($a_{\bf k}$) creates (annihilates) a boson with momentum ${\bf k}$.
$\epsilon_{\bf k}$  is obtained by taking the long wavelength limit of $\omega_{\bf k}$:
\begin{eqnarray}
\epsilon_{\bf k} = \frac{k_z^2}{2 m^*_{cc} }  + \frac{(k_x^2+k_y^2)}{2 m^*_{aa} }  .
\end{eqnarray}
with
\begin{equation}
\frac{1}{m^*_{\nu \nu}} = \frac{\partial^2 {\omega_{\bf k}}}
{{\partial k_{\nu}^2 }}\bigg|_{{\bf k=Q}} .
\end{equation}
The chemical potential $\mu$ is $g \mu_B H -\omega_{\bf Q} $. After a mean-field treatment of ${\cal H}_{\rm eff}$ in the PM phase, $H \leq H_{c1}$, the
interaction term simply leads to a renormalization of the chemical potential, $\mu \to {\tilde \mu }$,  with
\begin{equation}
{\tilde \mu }= \mu - 2 v_0 \rho
\label{self1}
\end{equation}
and the particle density
\begin{equation}
\rho = \frac{1}{N} \sum_{\bf k} \langle a^{\dagger}_{\bf k}  a^{\;}_{\bf k}  \rangle .
\label{self2}
\end{equation}
The resulting quadratic mean-field Hamiltonian can be easily diagonalized and the
various thermodynamic properties are computed by solving the self-consistent condition imposed by Eqs.\eqref{self1} and \eqref{self2}

The analytic calculations have been supplemented by large scale numerical
simulations of the microscopic model. We have used the Stochastic Series
Expansion (SSE) QMC method to simulate the Hamiltonian
(1) on finite-sized lattices using the experimentally determined parameters.
The SSE is a finite-temperature QMC technique based on importance sampling of
the diagonal matrix elements of the density matrix $e^{-\beta H}$. \cite{sse1,
sse2} The use of \textit{operator loop} cluster updates reduces the autocorrelation
time for the system sizes. We consider here up to $\approx 2\times 10^4$ spins
to at most a few Monte Carlo sweeps even at the critical temperature.\cite{dloops}
This enables us to explore the vicinity of the critical points very
efficiently. On the dense temperature grids needed to study the critical region
in detail, the statistics of the Monte Carlo results can be significantly
improved by the use of a parallel tempering scheme.\cite{tempering1,tempering2}
The implementation of this tempering scheme in the context of the SSE method has
been discussed in detail previously,\cite{ssetemp1,ssetemp2} and we follow
the one developed in Ref.\,[\onlinecite{ssetemp2}].

QMC estimates for observables of a spatially
anisotropic system can depend non-monotonically on the system size for isotropic
lattices. One can instead use anisotropic lattices to more
rapidly obtain monotonic behavior of the numerical results for extrapolating to
the thermodynamic limit. Anticipating similar effects in the present model
(since $J_a,J_b \ll J_c$), we have studied tetragonal lattices with
$L_x = L_y = L_z/4$.

The specific heat has been extracted from the simulation data by the numerical
differentiation of the total internal energy of the system -- a quantity that
is estimated extremely accurately by the SSE method. The temperature dependence
of the energy is approximated by a polynomial in $T$, and the derivative of the
polynomial fit is used to estimate the specific heat. Thus, artifacts of discrete numerical differentiation of the raw data are avoided and we yield a
relatively noise-free specific heat curve.

\section{Experimental Results}
\label{sec:exp}

\subsection{Magnetization}

\begin{figure}
\includegraphics[width=0.5\textwidth]{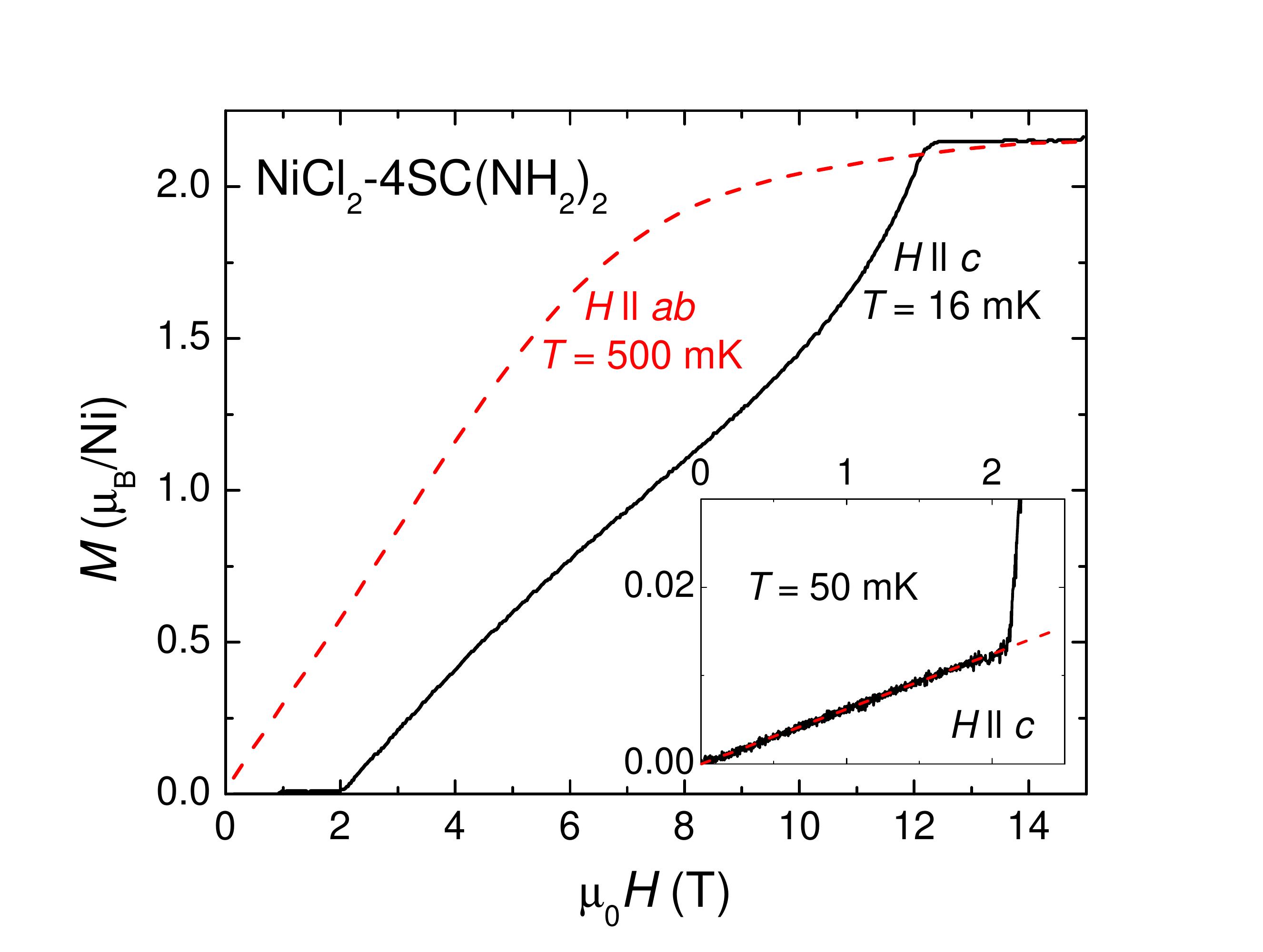}
\caption{(color online) Magnetization $M$ as a function of the magnetic field $H$ of DTN at 16\,mK for measurements $H$
parallel (solid line, Ref.\,[\onlinecite{PaduanFilho04a}]) and perpendicular to the crystallographic $c$ direction at 500\,mK (dashed line, ref.\,[\onlinecite{Zapf07}]). The inset shows new results of the magnetization at 50\,mK in low fields up to 2.5\,T and a linear fit (dashed line) to the data.}
\label{fig1_magn}
\end{figure}

\begin{figure}
\includegraphics[width=0.5\textwidth]{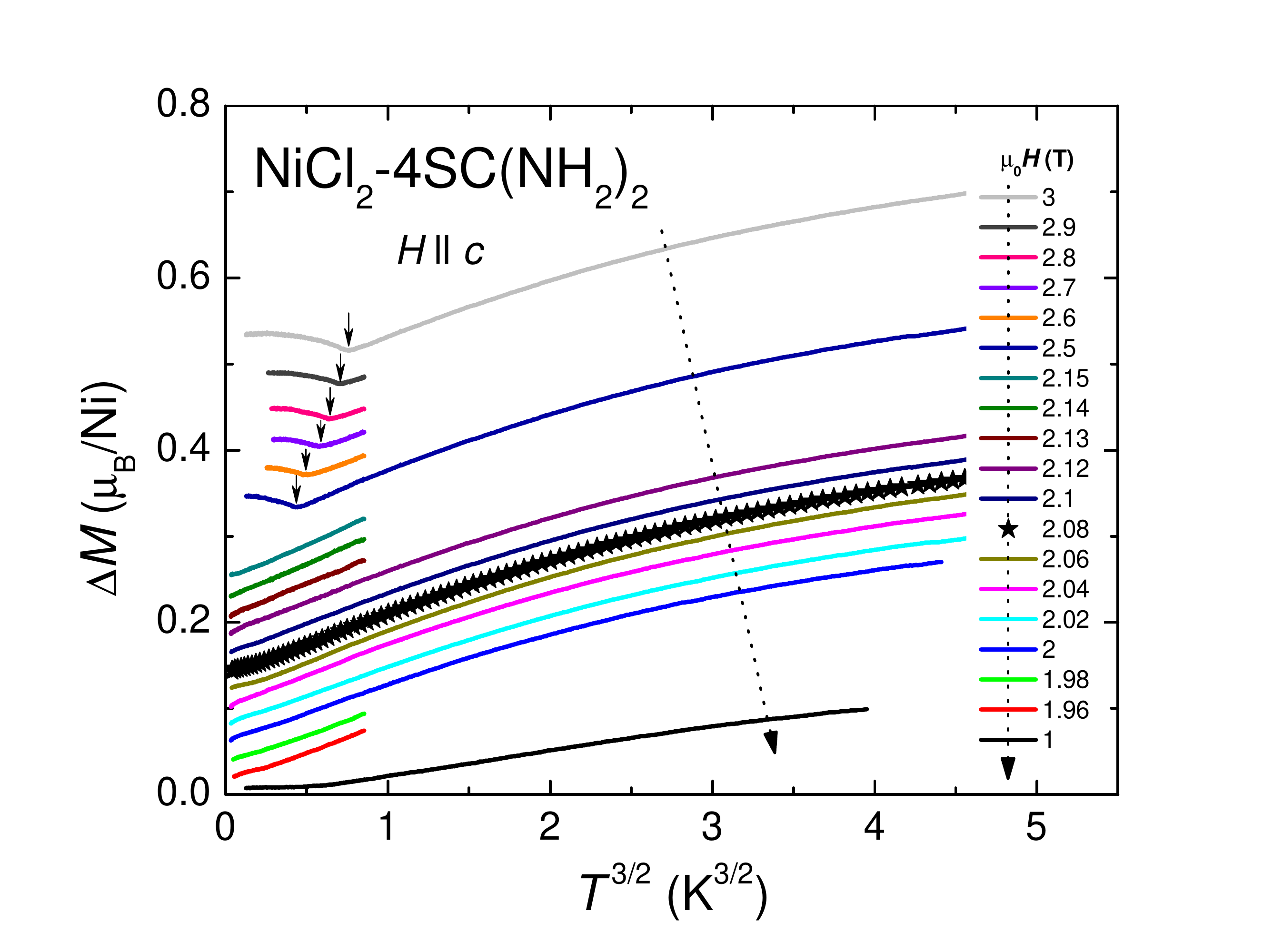}
\caption{(color online) Corrected magnetization $\Delta M$ versus $T^{3/2}$
for magnetic fields $H \parallel c$ between 1\,T and 3\,T. The magnetization at the critical field $\mu_{0}H_{c2}= 2.08$\,T is labeled with stars. Data above 1\,T are shifted vertically by $0.02\,\mu_{B}$ per Ni atom for better visualization. Arrows indicate the phase transition into the XY-AFM state.}
\label{fig2_magn}
\end{figure}

Figure\,\ref{fig1_magn} shows a comparison of the magnetization $M$ as a function of magnetic field measured perpendicular at $T=$\,500\,mK\cite{Zapf07} and parallel at $T=$\,16\,mK\cite{PaduanFilho04a} to the crystallographic $c$-axis. The latter data nicely reflect the N\'{e}el-ordered state in the $ab$ plane with increasing canting along $c$ between 2.1\,T and 12.6\,T followed by saturation. For $H \parallel ab$, the magnetization shows PM behavior with no ordering, approximately following a Brillouin function and
saturating at 2.2\,$\mu_{B}$ per Ni$^{2+}$-atom around 15\,T. In this field direction the magnetic field increases the size of the spin gap instead of closing it as happens for $H \parallel c$.
The inset of Fig.\,\ref{fig1_magn} shows the low field part of the magnetization $H \parallel c$ at 50\,mK in greater detail. We observe a linear increase between zero and the lower critical field $H_{c1}$, which can not be explained with a U(1) invariant Hamiltonian such as $\mathcal{H}$, where the magnetization is supposed to be zero at $T=0$ in the quantum PM region $H \leq H_{c1}$. This effect cannot be caused by single ion impurities, e.g. uncoupled Ni$^{2+}$ moments, because such spins should be fully polarized for magnetic fields well below the lower critical field $H \ll H_{c1}$. Instead, we conclude that the linear slope (dotted line) is caused by a misalignment of the sample, which gives a contribution $M(H \perp c)$ to the magnetization. From the value of the susceptibility compared to data for $H \perp c$ we estimate a misalignment of less than $1.2^{\circ}$.

Furthermore, we measured the magnetization for $H \parallel c$ near $H_{c1}$ and in the temperature range 0.1\,K$\leq\,T\,\leq $0.5\,K (data not shown) to extrapolate the phase boundary for $T \rightarrow$ 0.\cite{Yin08} We obtain a critical field $H_{c1}$ of 2.08\,T. It is important to note that this value depends on  the specific conditions under which the sample is mounted in the experimental setup. The slightly higher value of $H_{c1}$ in the magnetization compared to the specific heat and thermal expansion $L \parallel c$ values (see data below) supports the assumption of small sample misalignment, because angular resolved measurements of the magnetostriction
have shown that $H_{c1}$ increases with increasing angle between the field direction and the crystallographic $c$-axis.\cite{Zapf07}

The small PM contribution $M_{PM}$ to the magnetization made it necessary to refine the $M(T)$ data as a function of temperature for $H \parallel c$. Figure\,\ref{fig2_magn} shows the corrected values $\Delta M(T) = M - M_{PM}$ plotted versus $T^{3/2}$ for a variety of different magnetic fields 1\,T\,$< \mu_{0}H < $\,3\,T.
The magnetization is exponentially suppressed at low temperatures inside the quantum PM state ($\mu_{0} H = 1$\,T) and
develops a $T^{3/2}$-behavior when approaching the critical field $H_{c1}=2.08$\,T in agreement with the expected behavior for a BEC-QCP in 3 dimensions. The entrance into the XY-AFM state well above $H_{c1}$ is marked with a dip in the magnetization (arrows in Fig.\,\ref{fig2_magn}). Below the minimum inside the AFM phase, the magnetization increases with a power law for decreasing temperature.

\subsection{Specific Heat}

\begin{figure}
\includegraphics[width=0.5\textwidth]{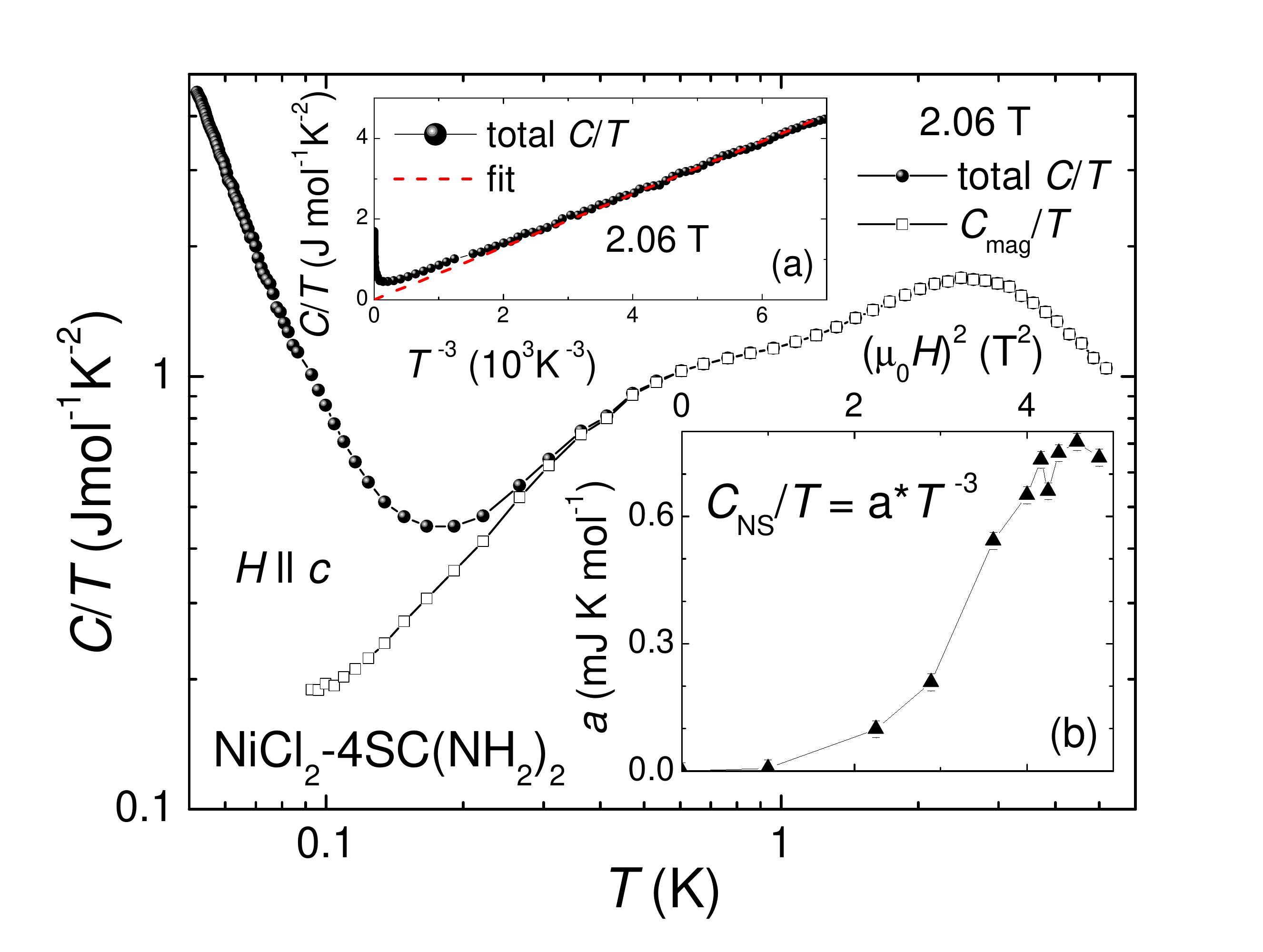}
\caption{(color online) Main panel: total specific heat $C/T$ as a function of temperature $T$ at the critical field $H_{c1} = 2.06$\,T (filled dots) and the resulting $C_{mag}/T$ (open squares) after subtraction of the nuclear Schottky (NS) contribution at low temperatures. Inset\,(a) shows the data of the main panel as $C/T$ versus $T^{-3}$ to illustrate the NS specific heat as indicated by a dashed line. The prefactor of the NS specific heat $C_{NS}/T=aT^{-3}$ is plotted in inset\,(b) versus $H^{2}$. It does not follow the expected $H^{2}$ field dependence.}
\label{fig3_Cp}
\end{figure}

We estimate the critical field $H_{c1}$ by MCE measurements (data not shown) following the analysis of the highest slope ($\partial T/\partial H$) of the temperature during field scans\cite{Zapf06} and find a value $H_{c1}=$\,2.06\,T in the specific-heat experimental setup.

The specific heat curve of DTN contains three contributions: nuclear Schottky (NS),  magnetic Schottky (MS) and  quantum-critical (QC) contributions. Each of them dominates in different regions of the $H-T$ phase diagram. The specific heat caused by phonons can be neglected in the temperature range below 5\,K. The NS contribution is difficult to master in this material, because it originates from several nuclei (H, N, Cl) generating a huge fraction of the specific heat at temperatures below 0.1\,K, see e.g. in Fig.\,\ref{fig3_Cp} the measurement at 2.06\,T. In addition,  the effective magnetic field (sum of the external field and the field generated by the ordered moments) becomes rather high in the AFM ordered state above $H_{c1}$. This increases the splitting of the nuclear energy levels further and the Schottky anomaly becomes very big. The inset\,(a) of Fig.\,\ref{fig3_Cp} demonstrates the way we subtracted the NS contribution from the original specific heat at the critical field $\mu_{0}H = 2.06\,T$. We fit the data between 80\,mK and 50\,mK as $C/T = a T^{-3}$, which is a good approximation for the high temperature behavior of the Schottky anomaly.\cite{Gopal66} The prefactor $a$ for the NS specific heat is given as a function of magnetic field in the inset\,(b) of Fig.\,\ref{fig3_Cp}. The pre-factors for different nuclei should be additive, $a = \sum a_{i}$, for the case that all the different nuclear energy levels are in the high temperature limit ($k_{B}T \gg \Delta_{nuclear} $). Because the Zeeman splitting is linear in magnetic field, $a(H)$ should obey a $H^2$ dependence for $H < H_{c1}$. This is not observed in DTN and presently not understood.

The insulating behavior of DTN is challenging for specific heat experiments, because no free electrons contribute to the thermal conductivity. Heat is carried only by magnetic excitations and  by phonons, whereas the phonon contribution is negligibly small in the temperature range below 1\,K. Therefore the different thermodynamic subsystems (nuclear spins, magnetic moments, lattice) are only poorly coupled to each other at low temperatures. This causes an out-of-equilibrium state of the sample at very low temperatures. Thus, we only include data above 80\,mK in our analysis.

\begin{figure}
\includegraphics[width=0.5\textwidth]{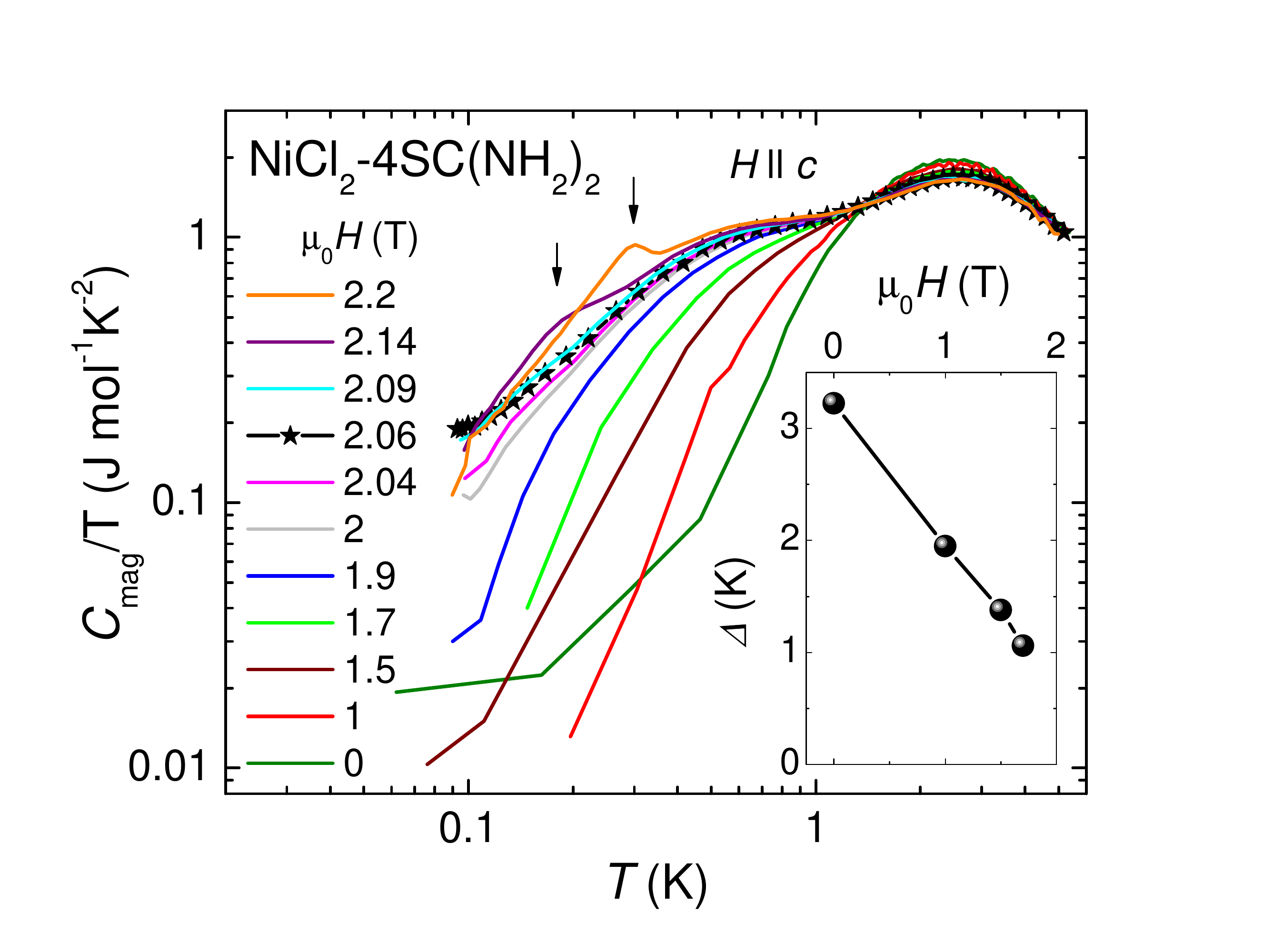}
\caption{(color online) Magnetic specific heat $C_{mag}/T$ as a function of temperature $T$ for magnetic fields between 0 and 2.2\,T in a double-log scale. The arrows indicate the AFM phase transition for fields $H>H_{c1}$. The critical field obtained from MCE experiments is 2.06\,T. The inset shows the energy gap between 0 and 1.7\,T, estimated from the exponential temperature increase of the magnetic specific heat at low temperatures.}
\label{fig4_Cp}
\end{figure}

Figure\,\ref{fig4_Cp} shows the magnetic specific heat $C_{mag}/T = C/T - C_{NS}/T$ in a double-logarithmic display after the subtraction of the nuclear Schottky contribution, $C_{NS}$, between 0 and 2.2\,T. The broad maximum around 2.5\,K in the zero field measurement is caused by the thermal population of the $| S^z=\pm 1>$ excited states that are a rather broad band due to dispersion caused by the exchange interactions.\cite{Zapf06} This dispersion is also responsible for the only slightly shift of the maximum to lower temperatures with higher magnetic fields. For small fields, the specific heat data can be fitted with an exponential function, $\exp (-\frac{\Delta}{k_{B}T})$ in the low temperature limit allowing us to extract the spin gap $\Delta$. The inset of Fig.\,\ref{fig4_Cp} shows the gap values, estimated from the experimental data. They decrease linearly from 3.22\,K down to 1\,K when $H$ varies between
$0$\,T and 1.7\,T, whereas the zero-field value is in close accordance with previous susceptibility results of 3.3\,K.\cite{PaduanFilho04b} Below 1.7\,T, the temperature range for exponential behavior is too small for reliable data fitting. From the zero field gap, $\Delta = 3.22$\,K, we can estimate the critical field,
\begin{equation}
\label{E_gap}
H_{c1} = \frac{ k_{B} \Delta}{g \mu_{B}},
\end{equation}
where the gap closes ($k_{B}=1.380\cdot 10^{-23}$JK$^{-1}$ and $\mu_{B}=9.274\cdot 10^{-24}$JT$^{-1}$).
The calculated value $H_{c1}=$\,2.12\,T matches the experimental values summarized in Table\,\ref{tab_Hc1} within 5\,\%.
The arrows in Fig.\,\ref{fig4_Cp} indicate the anomalies caused by the phase transition into the XY-AFM state in the 2.14\,T and 2.2\,T measurement.

\subsection{Thermal Expansion}

\begin{figure}
\includegraphics[width=0.5\textwidth]{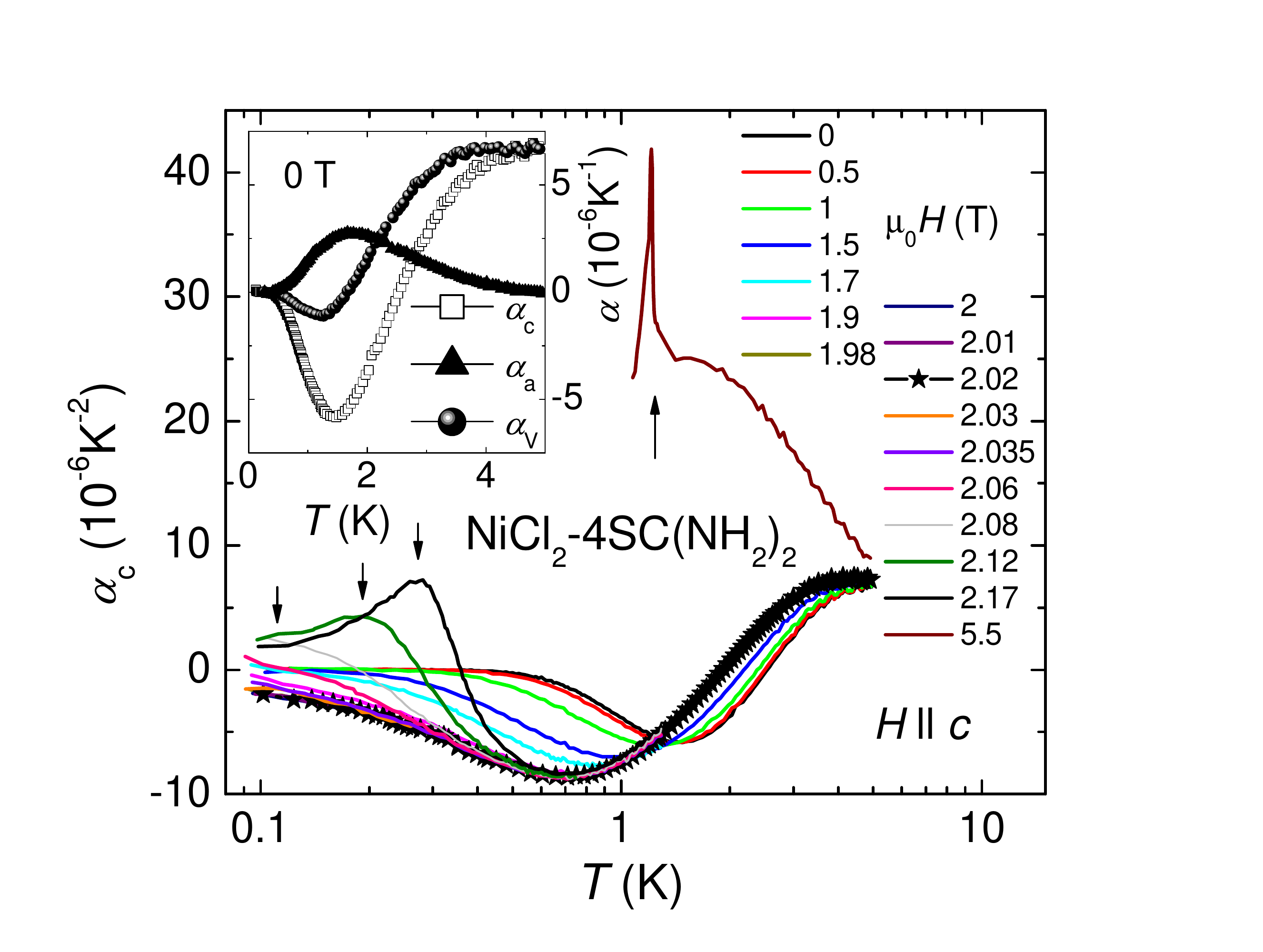}
\caption{(color online) Linear thermal expansion coefficient $\alpha_{c}$ measured along the crystallographic $c$-axis ($H \parallel c \parallel \Delta L_{c}$) as a function of the temperature $T$ between 0 and 5.5\,T, including the critical field $H_{c1}=2.02$\,T. Arrows mark the phase transition into the AFM ordered state. The inset displays in addition to $\alpha_{c}$, the coefficient $\alpha_{a}$ ($H \parallel c,\,\Delta L_{a}$) and the calculated volumetric coefficient $\alpha_{V}=2\alpha_{a}+\alpha_{c}$ for 0\,T.}
\label{fig5_alpha}
\end{figure}

The linear thermal expansion coefficient
\begin{gather} \label{linAlpha}
\alpha_{i} = \frac{1}{L_{0}} \frac{\partial \Delta L_{i}}{\partial T}
\end{gather}
is defined as the temperature derivative of the length change $\Delta L_{i}$ along a certain crystallographic direction $i$. In tetragonal systems, such as DTN, the volumetric expansion can be calculated from the linear coefficients
\begin{gather} \label{alpha_V}
\alpha_{V}=2\alpha_{a}+\alpha_{c}
\end{gather}
along the crystallographic $a$ and $c$ direction.

The main panel of Fig.\,\ref{fig5_alpha} shows $\alpha_{c}$ for $H \parallel c$ between 0 and 5.5\,T. For measurements well above $H_{c1}$, the transition into the ordered phase is indicated by a distinct anomaly, marked with arrows. In the temperature range up to 5\,K no significant contribution from the lattice is observed.
In zero field, $\alpha_{c}$ shows a rather broad minimum that shifts to lower temperatures in higher fields and can be attributed to the thermal population of energetically higher spin states $|S^{z}> = \pm\,1$, similar to the maximum in the specific heat. In zero field these spin states are equally occupied, because they  have same energy. The largest AFM exchange along the $c$-axis leads to the dominant magnetostrictive effect. Since the PM ground
state is a product of $S^z=0$ state to a good approximation, the thermal excitation of $S^z=\pm1$ states increases the nearest-neighbor XY AFM correlations along the $c$-axis
$\langle  {\bf S}_{\bf r} \cdot  {\bf S}_{{\bf r} +{\bf e}_c}  \rangle$. This increase leads to an attractive magnetostrictive force between nearest-neighbor ions along the $c$-axis that shrinks the lattice as the temperature
increases from zero. On the other hand, the magnetrostrictive force disappears at high enough temperature because $\langle  {\bf S}_{\bf r} \cdot  {\bf S}_{{\bf r} +{\bf e}_c}  \rangle \to 0$ for $T \to \infty $, implying that $\Delta L_c$ must have a minimum at a finite temperature where $\alpha_c$ changes sign. This expected behavior is fully consistent with the experimental results shown in Fig.\,\ref{fig5_alpha}.
The critical field for the thermal expansion measurements was estimated by detailed magnetostriction measurements, whereas the magnetostriction coefficient
\begin{gather} \label{linLambda}
\lambda_{i} = \frac{1}{\mu_{0}L_{0}} \frac{\partial \Delta L_{i}}{\partial H}.
\end{gather}
is definded as the magnetic field derivative of the length change along the $i$ direction. The analysis of the data (not shown) gives $H_{c1}=2.02$\,T for $\Delta L_{c}$ and $H_{c1}=2.08$\,T for $\Delta L_{a}$. The difference in the critical field values can be attributed to the application of small pressure on the sample during the experiment (see also section\,\ref{ssec:Ehrenfest}).

The inset of Fig.\,\ref{fig5_alpha} compares $\alpha_{c}$, $\alpha_{a}$ and $\alpha_{V}$ for the zero-field measurement. Between 5\,K and 3\,K, the thermal expansion is dominated by the length change along the $c$ direction, $\alpha_{c}>0$, $\alpha_{a} = 0$, because of the quasi-1 dimensional nature of the magnetic interactions in this temperature range.  In the temperature range below 3\,K, the thermal expansion coefficient $\alpha_{c}$ is negative and has the minimum that is expected because the curve $\alpha_{c}(T)$ must change sign at a finite temperature. In contrast, the thermal expansion coefficient in the plane $\alpha_{a}$ is positive for $T < 3\,K$ with a maximum at the temperature $T=T_{m}$ where the minimum occurs in $\alpha_{c}$. Calculating $\alpha_V$ via Equation\,(\ref{alpha_V}) reveals that the thermal expansion coefficients $\alpha_{a}$ and $\alpha_{c}$ strongly compensate each other and that the volume coefficient changes much less than the individual $\alpha_{i}$ values.

\section{Discussion}
\label{sec:dis}

\subsection{Critical Field $H_{c1}$}

\begin{table}
\begin{center}
\begin{tabular}{|c|c|} \hline
 Thermodynamic quantity & $H_{c1}$ (T)\\
 \hline \hline
 Magnetization $M(H,T)$ & 2.08 \\
 Magnetocaloric effect (MCE) & 2.06 \\
 Magnetostriction $\Delta L \parallel c$ & 2.02\\
 Magnetostriction $\Delta L \perp c$ & 2.08\\
 \hline
\end{tabular}
\end{center}
\caption{Critical field $H_{c1}$ estimated by different experimental methods and setups.
\label{tab_Hc1}}
\end{table}

The values for the lower critical field $H_{c1}$, estimated individually for the different experimental quantities, are summarized in Table \ref{tab_Hc1}. The values vary by $\pm\,30$\,mT around 2.05\,T. This difference can be attributed to the application of small pressure on the sample during the experiment. The spring-loaded capacitive dilatometer for thermal expansion and magnetostriction measurements can intrinsically apply a small amount of uniaxial pressure along the measured sample length. The misalignment of the crystal orientation out of $H \parallel c$ indicated by the magnetization experiments, is another reason for variations in $H_{c1}$.

\subsection{Comparison Theory - Experiment}

\begin{figure}
\includegraphics[width=0.5\textwidth]{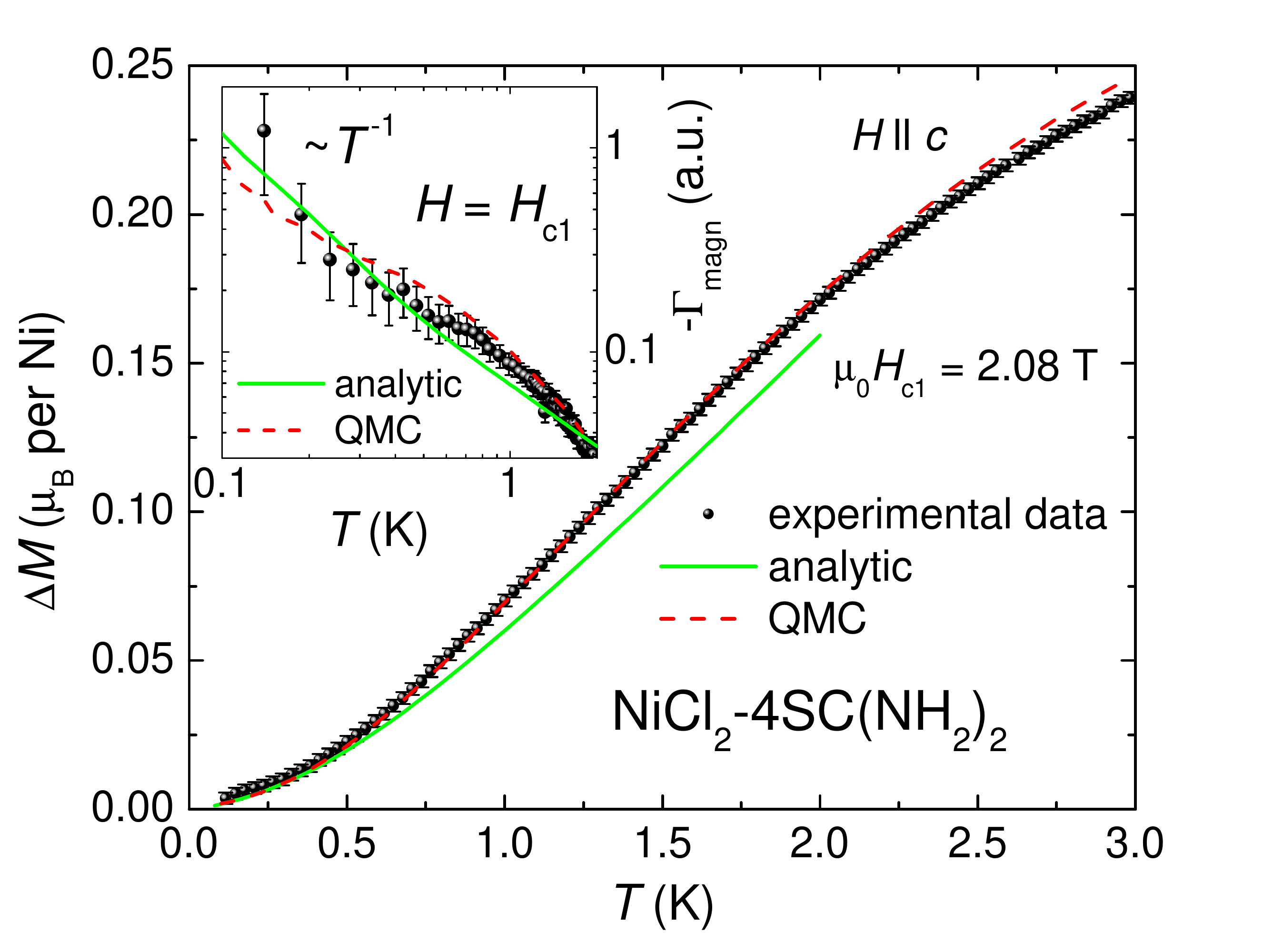}
\caption{(color online) Experimental magnetization $\Delta M(T)$ versus temperature $T$ (symbols) at the critical field $\mu_{0}H = 2.08$\,T applied along the crystallographic $c$ direction of DTN. The dashed line represents QMC results, whereas the solid line indicates the analytic calculations. The symbols in the inset show the magnetic Gr\"{u}neisen parameter $\Gamma_{mag}$ estimated from the data of the main panel and the specific heat values shown in Fig.\,\ref{fig4_Cp}. Dashed and solid lines represent QMC and analytical results respectively.
\label{fig6_magn}}
\end{figure}

Fig.\,\ref{fig6_magn} shows a comparison between the experimental magnetization (symbols) at the critical field and analytical calculations (solid line) and QMC simulations (dashed line). We observe that the QMC and analytic results agree with the experimental data within the error bars below 0.5\,K. A $T^{3/2}$ behavior is expected below 0.3\,K  for the BEC universality class, in contrast to $T^{2}$ dependence of an Ising-like QCP. While the analytic calculation is only valid at low temperatures (or low density of bosons) the QMC results remain valid at any temperature. This is the reason why the QMC results are in very good agreement
with the experimental data up to 2.2\,K.

\begin{figure}
\includegraphics[width=0.5\textwidth]{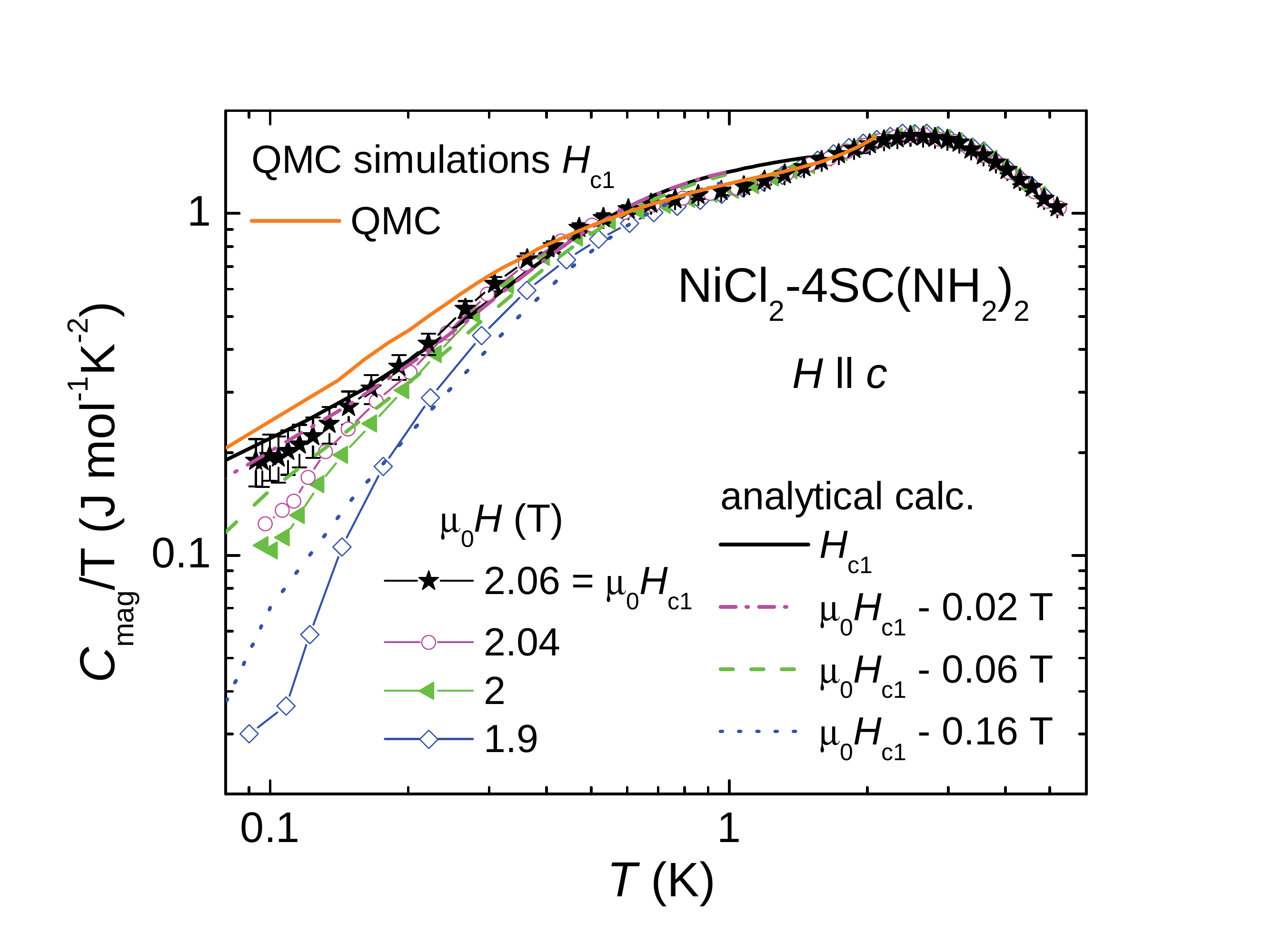}
\caption {(color online) Magnetic specific heat as $C_{mag}/T$ as a function of temperature $T$ at the critical field $H_{c1} = 2.06$\,T (stars), at 2.04\,T (open circles), 2\,T (filled triangles) and 1.9\,T (open diamonds) compared with QMC results at $H_{c1}$ (solid line) and with analytical calculations (dashed, dotted lines) at $H_{c1}$ and for fields below the critical field.}
\label{fig7_Cp}
\end{figure}

Figure\,\ref{fig7_Cp} shows the experimental data of the specific heat, $C_{mag}/T(T)$, (symbols)
at and slightly below the critical field $H_{c1}$, compared with data of analytic
calculations
(broken lines) and QMC simulations (solid line) in a double logarithmic display.
The experimental data  exhibit a $\sqrt{T}$ dependence in the low temperature
limit at the critical field -- in agreement with mean-field
calculations and  QMC simulations --  which is characteristic for the 3-dimensional BEC universality class.
We observe that the experimental data agree well within error bars with the analytic results
down to the lowest temperatures, whereas there is a slight deviation of the QMC results below 0.3 K. This discrepancy is most likely due to a small error
in the numerical determination of the critical field $H_{c1}$.
A linear temperature dependence of the specific heat
$C_{mag}/T(T)$ is expected for an Ising-like QCP, which we can exclude from our measurements. The broad Schottky
maximum around 2.5\,K in the experimental data originating from the population of
$|S^{z}> = \pm\,1$ excited spin states, can only be reproduced by the QMC simulations because they remain valid up to arbitrarily high temperatures.
Seen in the experimental data as well as in the QMC results is the crossover from 3 dimensional ($C/T \sim T^{1/2}$) to 1 dimensional ($C/T \sim T^{-1/2}$) behavior of the specific heat at higher temperatures for the measurement at $H_{c1}$. The change of slope is marked with a broad hump around 0.5\,K. Note that the 1-dimensional temperature dependence overlaps with the above mentioned Schottky contribution to the specific heat.

For $H < H_{c1}$ measurements, the analytical curves as well as the
experimental values lie below the specific heat data at the critical field.
They grow smaller as the distance from $H_{c1}$ increases. This observation
confirms the correct estimation of the value $H_{c1} = 2.06$\,T for the critical field in the
MCE measurements. Furthermore, Fig.\,\ref{fig7_Cp} shows nice agreement between
analytical predictions and experimental data at low temperatures for all shown fields.
Deviations seem to be larger for $H < H_{c1}$, but this is an effect of the
double-logarithmic display.

In presence of Ising-like anisotropy, the gap should reopen inside the AFM phase. We do not observe any exponential
temperature dependence in the measurements for fields above $H_{c1}$ in Fig.\,\ref{fig4_Cp},
namely 2.09\,T, 2.14\, and 2.2\,T. This, however, could also be due to (i) lack of data
at temperatures below 80\,mK and (ii) the onset of the phase transition seen as a
broad anomaly in the specific heat.

\begin{figure}
\includegraphics[width=0.5\textwidth]{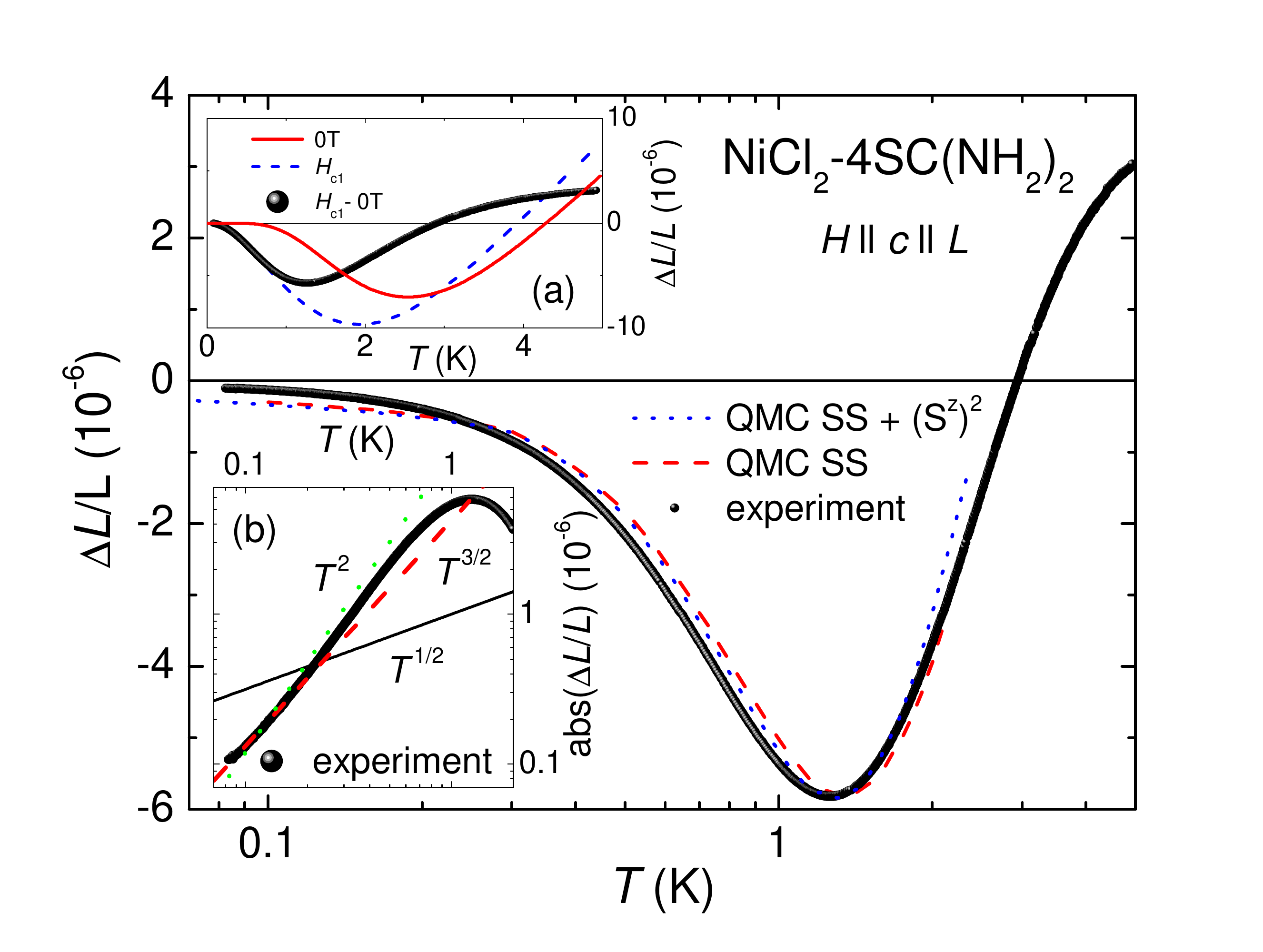}
\caption{(color online) The main panel shows the normalized length change $\Delta L/L$ (symbols) as a function of the temperature $T$ of DTN for $H \parallel c \parallel L_{c}$ at the critical field $H_{c1}=2.02$\,T in a semi-logarithmic display. The experimental data are compared with QMC calculations, where the spin-spin correlator (SSC) (dashed line) and the SSC together with the $(S^{z})^{2}$ term (dotted line) are taken into account. The inset\,(a) shows the experimental data at zero field (solid), the critical field $H_{c1}=2.02$\,T (dotted line) and the difference $\Delta L_{c}/L(H_{c1})-\Delta L_{c}/L(0))$ (symbols). In inset\,(b) the absolute values $\Delta L/L$ (symbols) are plotted as a function of $T$ in comparison with $\sim T^{1/2}$ (solid), $\sim T^{3/2}$ (dashed), and $\sim T^{2}$ (dotted line) temperature dependence.
\label{fig8_L}}
\end{figure}

Before we discuss the scaling behavior of the thermal expansion coefficient at the
critical field, let us have a closer look at the length change $\Delta L_{c}/L$,
which is shown in Fig.\,\ref{fig8_L} along with results from QMC simulations. The
inset\,(a) of Fig.\,\ref{fig8_L} shows the experimental length at the critical
field after the subtraction of the data in 0\,T, in order to separate quantum
critical from the non-critical magnetic contributions of the sample. The same
procedure was done for QMC data. In QMC simulations, $\Delta L_{c}/L$ at $H_{c1}$
is obtained from the estimation of the spin-spin correlator (SSC),
$\langle {\bf S}_{\bf r} \cdot {\bf S}_{\bf r +{\bf e}_{\nu}}\rangle $\cite{Zapf08} and optionally
additional terms.
The main panel of Fig.\,\ref{fig8_L} shows the comparison between the experimental
data and QMC simulations. The qualitative features are well reproduced by the expectation
value of the SSC for temperatures above 0.3\,K.
The scaling factor between experiment
and theory is $1.85\cdot 10^{-4}$.
$\Delta L_{c}/L$  follows a $T^{\gamma}$ power law with $\gamma$ between 2 and 3/2 as illustrated in inset\,(b) of Fig.\,\ref{fig8_L}.
This result is in close accordance with the expected BEC behavior of $\sim T^{3/2}$
The discrepancy between experimental data and QMC simulations below 0.3\,K can be attributed to additional contributions to $\Delta L_{c}/L$ besides the SSC. However, an additional consideration of a $\langle (S_i^z)^2 \rangle$ term that is expected from symmetry arguments, does not improve the agreement significantly. Therefore, the origin of this discrepancy remains unclear at the present level of analysis.

\begin{figure}
\includegraphics[width=0.5\textwidth]{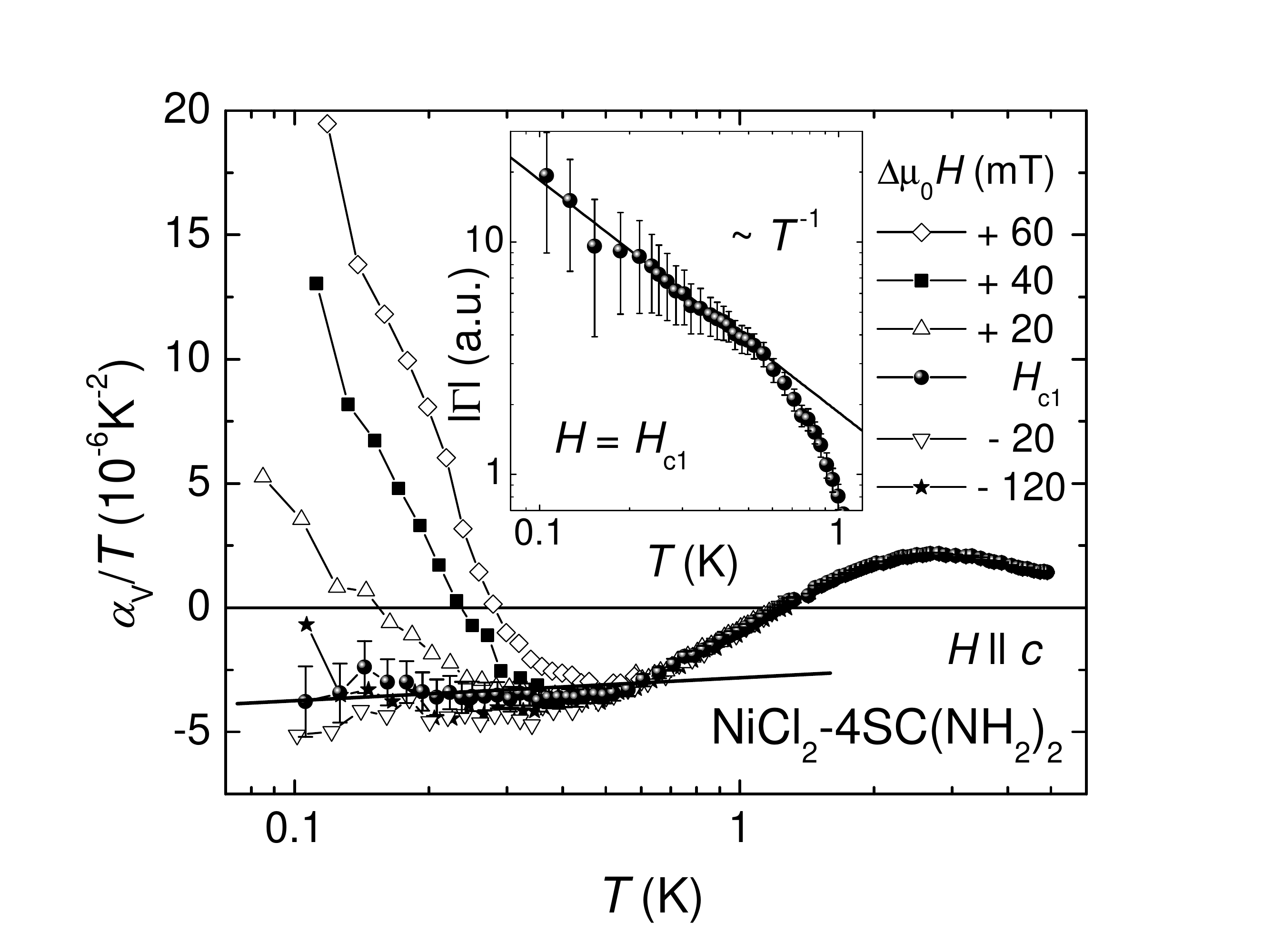}
\caption{The volume thermal expansion coefficient divided by $T$ is presented for fields close to the critical fields as a function of temperature $T$ in a semi-logarithmic plot for DTN with field direction $H \parallel c$. The inset shows the absolute value of the thermal Gr\"{u}neisen parameter $\Gamma_{th} = \alpha_{V}/C$ at the critical field $H_{c1}$. The solid line in the inset illustrates $T^{-1}$ behavior.
\label{fig9_thG}}
\end{figure}

Figure\,\ref{fig9_thG} summarizes the volumetric thermal expansion coefficient divided by the temperature $\alpha_{V}/T$ for fields at and close to the QCP. Note, that according to equation\,(\ref{alpha_V}) different critical values $H_{c1}$ were considered for $\alpha_{a}$ and $\alpha_{c}$ and taken into account for the estimation of $\alpha_{V}$.
We find, that at $H_{c1}$ the values $\alpha_{V}/T$ have a weak, but finite temperature dependence (solid line) meaning that the thermal expansion coefficient $\alpha_{V}$ diverges as expected at the QCP. For magnetic fields $H \leq H_{c1}$ the low-temperature values $\alpha_V/T$ show similar behavior.

In general, pressure $p$ and magnetic field $H$ are equivalent parameters of the free energy $F(p,H)$ in close vicinity to the critical field $|H_{c1}-H|\ll H_{c1}$. Therefore, thermodynamic quantities derived from pressure and field dependencies can be converted
\begin{gather} \label{convert}
\frac{\partial}{\partial p} = \Omega \frac{\partial}{\partial H}
\end{gather}
by multiplication with the prefactor $\Omega = \partial H_{c1}/ \partial p$, which is the hydrostatic pressure dependence of the critical field $H_{c1}$.

Equation\,(\ref{convert}) implies that the compressibility $\kappa = \partial^{2}F/\partial p^{2}$ is proportional to the magnetic susceptibility that is a step function of the magnetic field at $H_{c1}$ and is shown in Fig.\,\ref{fig10_lambda_chi}. Therefore, $\kappa$ increases rapidly at $H_{c1}$ and leads to a softening of the crystal lattice, recently demonstrated experimentally on DTN by ultrasound experiments.\cite{Chiatti08} Considering this, we speculate, that the huge change in the lattice properties is responsible for the deviation of $\alpha_{V}$ from the exact behavior of a BEC. Similar crystal softening as a precursor for quantum criticality was e.g. also observed at the metamagnetic transition in CeRu$_{2}$Si$_{2}$.\cite{Weickert10}

The thermal expansion coefficient $\alpha_{V}$ for fields above $H_{c1}$ shows the clear onset of the phase transition with positive values $\alpha_{V} > 0$.

\subsection{Gr\"{u}neisen Parameter}

The magnetic Gr\"{u}neisen parameter, $\Gamma_{mag}$, is given in the inset of Fig.\,\ref{fig6_magn} and compared with theoretical data of the QMC simulation and the analytical calculations.
In the temperature range below 0.3\,K, the experimental $\Gamma_{mag}$ shows the onset of divergence, as expected for a QCP. As far as it is observable in the low temperature limit, $\Gamma_{mag}$ follows the theoretical prediction $\sim T^{-1}$, because the magnetization and specific heat obey the expected behavior individually. For the thermal Gr\"{u}neisen parameter, $\Gamma_{th}$, shown in the inset of Fig.\,\ref{fig9_thG}, a $T^{-1}$ behavior is found as well for temperatures below 0.6\,K. These observations are in fully agreement with the universality class of a BEC QCP.


\subsection{Ehrenfest relations}
\label{ssec:Ehrenfest}

\begin{figure}
\includegraphics[width=0.5\textwidth]{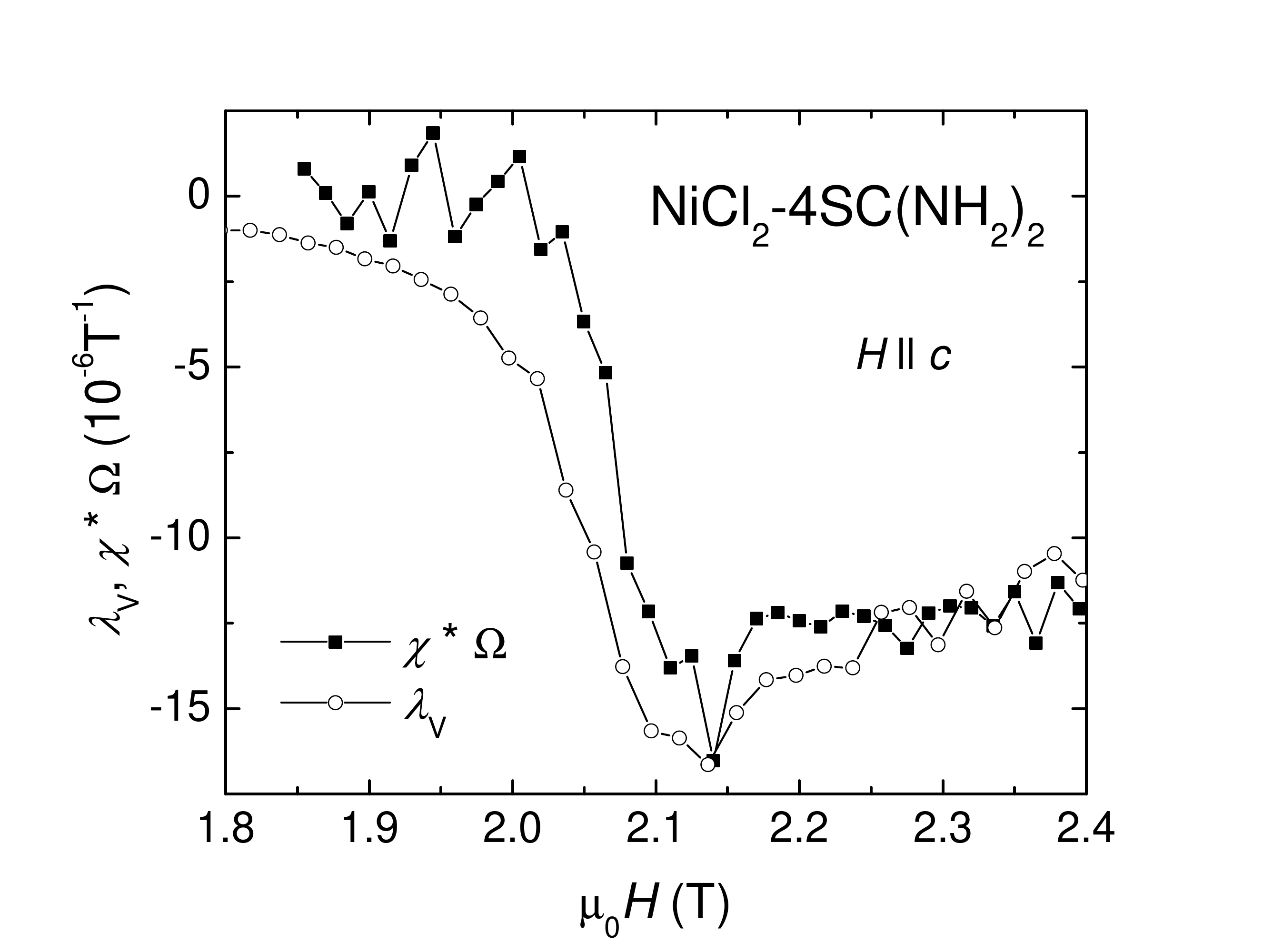}
\caption{Volume magnetostriction coefficient $\lambda_{V}$ and susceptibility $\chi = \mu_{0}^{-1}\partial M /\partial H$ at 0.1\,K of DTN. The susceptibility is scaled by the hydrostatic pressure dependence of the critical field $\Omega$ estimated from the Ehrenfest relations.}
\label{fig10_lambda_chi}
\end{figure}

The anomalies at the phase boundaries observed in the thermal expansion coefficient, $\alpha_{i}$, the specific heat, $C$, the magnetostriction coefficient, $\lambda_{i}$,
and the magnetization, $M$, allow us to obtain the pressure dependence of (i) the transition temperature $T_{N}$
\begin{gather} \label{Ehrenfest_T}
\frac{\partial T_{N}}{\partial p_{i}} = V_{m} T_{N}\frac{\Delta \alpha_{i}}{\Delta C_{p}}
\end{gather}
and of (ii) the critical field $H_{c1}$
\begin{gather} \label{Ehrenfest_H}
\frac{\partial H_{c1}}{\partial p_{i}} = V_{m} \frac{\Delta \lambda_{i}}{\Delta \chi}
\end{gather}
by the Ehrenfest relations, which hold at phase transitions of second order. Equations\,(\ref{Ehrenfest_T},\ref{Ehrenfest_H}) are valid for uniaxial as well as hydrostatic pressure, dependent if the linear or volume coefficients $\lambda_{i}$ or $\alpha_{i}$ are used in the analysis.

We find $\partial T_{N}/\partial p_{c} = 18.5$\,K\,GPa$^{-1}$ at $H = 2.2$\,T for uniaxial pressure applied along the crystallographic $c$ direction. This value is rather high, even compared to \TlCu , which already exhibits a huge value of several Kelvin per GPa dependent on the considered field range.\cite{Johannsen05} Direct measurements of the thermal expansion $\Delta L_{a}$ are currently not available, however, from measurements in small magnetic fields $H<H_{c1}$, we expect a negative uniaxial pressure dependence $\partial T_{N}/\partial p_{a}$  that should partially cancel the value (18.5\,K\,GPa$^{-1}$) along $c$, leading to a smaller but positive hydrostatic value. Similar behavior was observed in \TlCu .\cite{Johannsen05}

The pressure dependence of the critical field, $\partial H_{c1}/\partial p_{i}$, could be obtained at 0.1\,K for both directions $c$ and $a$. Because of a negative jump $\Delta \lambda_{c}$ in the magnetostriction coefficient at $H_{c1}$ (data not shown), the value $\partial H_{c1}/\partial p_{c}$= -\,6.76\,T\,GPa$^{-1}$ is negative. In contrast, $\partial H_{c1}/\partial p_{a}$=1.84\,T\,GPa$^{-1}$ is positive and the resulting hydrostatic pressure dependence, which is equivalent to the prefactor $\Omega$ in equation\,\ref{convert}, adds to -\,3.07\,T\,GPa$^{-1}$. The correct estimate of $\Omega$ can be proved by a comparison of $\lambda_{V}$ with the magnetic susceptibility $\chi$ measured at 0.1\,K. Both thermodynamic quantities are linked via equation\,(\ref{convert}) and it follows
\begin{gather} \label{link}
\lambda_{V} = \frac{\partial^{2}F}{\partial p \,\partial H} = \Omega \,\frac{\partial^{2}F}{\partial H^{2}} = \Omega\,\chi.
\end{gather}
Figure\,\ref{fig10_lambda_chi} shows an excellent agreement between the magnetostriction coefficient $\lambda_{V}$ and the scaled susceptibility $\Omega \chi$. We conclude from the analysis of the Ehrenfest relations, that the application of uniaxial pressure $p_{c}$ along the $c$ direction increases the ordering temperature $T_{N}$ and  reduces the critical field $H_{c1}$. The behavior is opposite for uniaxial pressure
along the $a$ direction. The response of DTN to hydrostatic pressure is dominated by the uniaxial pressure dependence along the $c$ axis because this is the direction of the dominant magnetic exchange interaction.


\section{Summary }
\label{sec:sum}

We present a comprehensive experimental and theoretical study of the thermodynamic properties: specific heat, magnetization, and thermal expansion in the vicinity of the field-induced QCP around $H_{c1} \approx 2$\,T in \DTN . This point marks the entrance into an 3-dimensional XY antiferromagnetically ordered state and can be described within the formalism of a BEC of magnons. We find  a $T^{3/2}$ low temperature behavior of the specific heat and the magnetization at $H_{c1}$ that are in agreement with the universality class of a BEC-QCP. The thermal expansion coefficient shows a temperature dependence $T^{\gamma}$ with $3/2 < \gamma < 2$ for $T \rightarrow 0$, which is in close agreement with expectations for this kind of QCP. QMC simulations nicely reproduce the features observed in the magnetization, specific heat and thermal expansion over a broad temperature range. Only the low temperature dependence in the specific heat and thermal expansion deviates due to intrinsic uncertainties in the simulation method. Furthermore, we analysed the thermal, $\Gamma_{th}$, as well as the magnetic Gr\"{u}neisen parameter $\Gamma_{mag}$, which are key quantities for the identification of QCPs and diverge with specific power laws. Experimentally, we found a $T^{-1}$
divergence for $\Gamma_{mag}$ and $\Gamma_{th}$ as expected for a BEC-QCP. Moreover, we estimated the influence of pressure on the transition temperature, $T_{N}$, and the critical field, $H_{c1}$ and found opposite effects for uniaxial pressure along the crystallographic $a$ and $c$ axes. Uniaxial pressure along $c$ ($a$) direction increases (reduces) the ordering temperature $T_{N}$ while it reduces (increases) the critical field $H_{c1}$. Due to the quasi-1 dimensional character of the exchange interactions in DTN, the application of hydrostatic pressure increases $T_{N}$ and reduces $H_{c1}$. Our results encourage pressure experiments that shift $H_{c1}$ to zero field. Since the field induced QCP is BEC-like, the pressure induced QCP should belong to the $O(2)$ universality class in dimension $D=3+2$.
\\
\acknowledgments
FW was funded by the MPG Research initiative: \textit{Materials Science and Condensed Matter Research at the Hochfeldmagnetlabor Dresden}.
MJ acknowledges hospitality at the MPI for Chemical Physics of Solids, where the experiments were carried out. VSZ acknowledges funding via LDRD/DR project 20100043DR and LP was partially supported by CONACyT.

\appendix

\end{document}